\documentclass[useAMS,usenatbib]{mn2e}
\usepackage{epsfig}
\usepackage{natbib}
\title{Compact Stellar Systems around NGC 1399}
\author[Firth et al.]
       {\parbox{16cm}{P. Firth\thanks{E-mail: firth@physics.uq.edu.au}$^{1}$, M. J. Drinkwater$^{1}$, E.A. Evstigneeva$^{1}$, M. D. Gregg$^{2}$, A. Karick$^{2}$, J. B. Jones$^{3}$ and S. Phillipps$^{4}$}\\
        \\
       ${1}$ Department of Physics, University of Queensland, QLD 4072, Australia.\\
       ${2}$ Physics Dept, U.C. Davis, and IGPP, Lawrence Livermore National Laboratory, L-413,Livermore, CA 94550, USA.\\
       ${3}$ Astronomy Unit, School of Mathematical Sciences, Queen Mary University of London, Mile End Road, London E1 4NS, United Kingdom.\\
       ${4}$ Astrophysics Group, H.H. Wills Physics Laboratory, University of Bristol, Tyndall Avenue, Bristol BS8 1TL, United Kingdom.}
\date{Accepted 2007 August 1, Received 2007 July 1}
\pagerange{\pageref{firstpage}--\pageref{lastpage}} 
\pubyear{2007}

\begin{document}

\bibliographystyle{mn2e}

\maketitle

\label{firstpage}

\begin{abstract}
We have obtained spectroscopic redshifts of colour-selected point sources in four wide area VLT-FLAMES fields around the Fornax Cluster giant elliptical galaxy NGC 1399, identifying as cluster members 30 previously unknown faint ($-10.5<M_{g^\prime}<-8.8$) compact stellar systems (CSS), and improving redshift accuracy for 23 previously catalogued CSS.

By amalgamating our results with CSS from previous 2dF observations and excluding CSS dynamically associated with prominent (non-dwarf) galaxies surrounding NGC 1399, we have isolated 80 `unbound' systems that are either part of NGC 1399's globular cluster (GC) system or intracluster GCs. For these unbound systems, we find (i) they are mostly located off the main stellar locus in colour-colour space; (ii) their projected distribution about NGC 1399 is anisotropic, following the Fornax Cluster galaxy distribution, and there is weak evidence for group rotation about NGC 1399; (iii) their completeness-adjusted radial surface density profile has a slope similar to that of NGC 1399's inner GC system; (iv) their mean heliocentric recessional velocity is between that of NGC 1399's inner GCs and that of the surrounding dwarf galaxies, but their velocity dispersion is significantly lower; (v) bright CSS ($M_V<-11$) are slightly redder than the fainter systems, suggesting they have higher metallicity; (vi) CSS show no significant trend in $g^\prime - i^\prime$ colour index with radial distance from NGC 1399.
\end{abstract}


\section{Introduction}
% Cosmological context of $\Lambda$CDM and reionisation, missing satellite problem, hypothesised radial density of CSS in clusters. Why are our observations important?

The $\Lambda$CDM model of structure formation predicts many more satellite dark matter concentrations (sub-haloes) than are observed as light from their embedded baryonic matter -- this is the well-known missing satellite problem \citep*[e.g.][]{Kravtsov..2004}. Several explanations address this mismatch between theory and observation - for example that the $\Lambda$CDM model needs correction, that the dark matter dominated satellites are too faint to detect with present instruments, or that star formation was obstructed in the smaller dark matter concentrations during cosmic reionisation \citep{Moore..2006}. Recent progress in this field has included the creation of increasingly high resolution computer simulations and the detection in our Local Group of faint, dark matter dominated stellar systems \citep{Belokurov..2007}.

Nearby galaxy clusters are physically and observationally appropriate environments to assess the frequency and distribution of compact stellar systems (CSS), the most luminous being potential sites of small-scale dark matter concentrations. These systems range in size from globular clusters (GCs: not considered to have dark matter) to dwarf galaxies (likely to have dark matter), and the intermediate-sized ultra-compact dwarf galaxies (UCDs: which may have dark matter). This paper reports a further step in our ongoing investigation of the distribution of CSS within the Fornax galaxy cluster.

The densely populated and highly evolved gravitational centres of galaxy clusters are ideal locations to study the interaction, merging and destruction of galaxies and sub-galactic stellar systems. GCs and other CSS surrounding central cluster galaxies provide observable fossil evidence of cluster evolution \citep[e.g.][]{Moore..1999}. In nearby galaxy clusters, CSS are attractive observing targets as potential trace particles of galaxy assembly \citep[e.g.][]{Blakeslee..2000, Brodie..2006} -- for example GCs, numerous around giant elliptical galaxies, are resistant (particularly the most massive examples) over long periods to disruption through tidal forces, and their metal abundance features confirm that they are as old as their host galaxies \citep{Carretta..2000}. Being relatively bright point sources, CSS are excellent targets for wide-field photometry and multi-fibre spectroscopy.

Three major sub-populations of CSS are discussed in the literature.

\begin{itemize}
 \item \textbf{Globular clusters (GCs)} are gravitationally bound to a specific host galaxy. These are normally found at $M_V>-11$ ($V \sim20.4$ at the distance of the Fornax Cluster), and this magnitude marks a break in the metallicity and size distribution of Fornax CSS according to \citet{Mieske..2006I}. Excluding the light-saturated innermost region, NGC 1399's GC population has been progressively catalogued by several authors -- \citet{Dirsch..2004} published a spectroscopically-confirmed inner-GC system catalogue of 468 GCs (2 to 9 arcminutes from NGC 1399), and \citet{Schuberth..2004} added 160 GCs in the outer parts of the system (8 to 18 arcminutes from NGC 1399, although individual positions were not published). Due to incomplete wide-area redshift data for faint point source targets, the radial extent of NGC 1399's GC system is uncertain -- but it is estimated at $45\pm5$ arcminutes (220 to 275 kpc) from wide-area photometry \citep{Bassino..2006a}.
 
 \item \textbf{Intracluster globular clusters (IGCs)} are a proposed population of GCs that are not gravitationally bound to a specific host galaxy, having either been tidally stripped from their host galaxy or formed \textit{in situ} \citep{West..1995, Bekki..2003II}. \citet{West..1995} proposed that this potentially large population of IGCs might account for the high GC specific frequency ($S_N$) of cluster core galaxies such as NGC 1399. IGCs are identified observationally by their location in intracluster space, outside the accepted radial and recessional velocity limits of galaxy-associated GC systems. Wide-area photometry by \citet{Bassino..2003} suggested candidate IGCs up to 160 arcminutes from NGC 1399 which are not associated with other galaxies, and recently published wide-area spectroscopy \citep{Bergond..2007} has confirmed 61 isolated IGCs up to 34 arcminutes from NGC 1399. However, since cluster core giant elliptical galaxies have very extended GC populations and we lack information on their true spatial motion, it is practically impossible to conclusively determine whether specific GCs are really IGCs.
 
 \item \textbf{Ultra-compact dwarfs (UCDs)} are massive and relatively bright CSS with absolute magnitudes $-13 \leq M_V \leq -11$ and typical radii 3 to 5 times larger than GCs. Alternatively labelled ultra-compact objects or dwarf-globular transition objects (\citealt*{Mieske..2002}; \citealt{Hasegan..2005}), they extend into the fundamental plane gap between GCs and dwarf galaxies. Three origins have been proposed for these systems -- that they are the bright tail of the GC distribution \citep[e.g.][]{Hilker..1999a, Drinkwater..2000b, Phillipps..2001, Mieske..2002}; or massive superstellar clusters formed in galaxy interactions \citep{Fellhauer..2002, Maraston..2004}; or the tidally-stripped remnants of early-type nucleated dwarf galaxies (\citealt*{Bekki..2001}; \citealt{Bekki..2003I}). Following the detection of two unusually bright but compact objects by \citet{Hilker..1999a}, 60 spectroscopically-confirmed `UCDs' (both bright and faint) have been catalogued in the central region of the Fornax Cluster \citep{Drinkwater..2000b, Drinkwater..2004, Gregg..2007}.
\end{itemize} 

We will avoid the above descriptors, since they assume information about kinematics and origins which we do not possess. Instead we identify CSS sub-populations from an observational perspective as follows: (1) \textit{bound CSS} are gravitationally bound to galaxies surrounding NGC 1399; the remaining CSS are either (2) \textit{bright unbound CSS} with $M_V<-11$ \citep{Mieske..2006I} or (3) \textit{faint unbound CSS} with $M_V\geq-11$.

In this paper we describe 2004 November observations with the VLT-FLAMES multi-object spectrograph of the central region of the Fornax Cluster, and investigate the distribution of CSS around the giant elliptical galaxy NGC 1399. Our aims are twofold -- to extend the redshift-confirmed catalogue of radially-dispersed CSS surrounding NGC 1399; and to investigate whether there are kinematically and photometrically distinct sub-populations \citep{Mieske..2006I} of bright and faint CSS that are not gravitationally bound to other cluster galaxies.

In section 2 we describe our VLT-FLAMES observations and results. In section 3 we separate the CSS likely to be gravitationally bound to prominent galaxies surrounding NGC 1399. In section 4 we analyse the projected radial distribution, recessional velocity dispersion and photometric properties of the unbound CSS, and in section 5 we summarise our key findings.

Throughout this paper we assume a distance to NGC 1399 of 18.6 Mpc \citep{Richtler..2000} which corresponds to a distance modulus ($m-M$) $\simeq 31.35$.

\section{VLT-FLAMES Observations}
The region within a $1^\circ$ ($\sim350 \mbox{kpc}$) radius of NGC 1399 contains numerous extended light sources (mostly elliptical and lenticular galaxies) and unresolved point sources. Photometric catalogues of this region (\citealt{Ferguson..1989}; \citealt{Hilker..1999a}; \citealt{Hilker..1999b}; \citealt*{Karick..2003}) provide a good starting point to study the distribution of CSS, since the numerous GCs around NGC 1399 dominate the point source population (compared to foreground Galactic stars and background galaxies) and can be fairly reliably colour selected.

In order to unambiguously eliminate foreground stars and background galaxies and to locate individual cluster members, we have obtained spectroscopic redshifts of colour-selected point sources over a relatively large projected area (four 25-arcminute diameter fields surrounding NGC 1399). Fortunately, the Fornax Cluster is relatively isolated in redshift space -- its recessional velocity boundaries are 600 to 2500 $\mbox{km} \,\mbox{s}^{-1}$ \citep{Phillipps..2001, Gregg..2007}. Our results add to those from the original compact object redshift surveys by \citet{Hilker..1999a} and \citet{Drinkwater..2000a}, and subsequent deeper redshift surveys \citep{Dirsch..2004, Drinkwater..2004, Schuberth..2004, Bergond..2007, Gregg..2007}.

We previously surveyed the region around NGC 1399 for CSS with the Two-degree Field multi-object spectrograph (2dF) on the Anglo-Australian Telescope (AAT) down to $b_J\sim$19.8 \citep{Drinkwater..2000b}, and the work of \citet{Gregg..2007} extends this to $b_J\sim$21.5. Our VLT-FLAMES observations were designed to access fainter targets ($r^\prime < 22.75$ or approximately $b_J < 23.25$) with higher-resolution spectroscopy. Although we did not achieve signal-to-noise levels sufficient to study internal stellar populations, we have extended the faint end of the wide-area CSS distribution in the Fornax Cluster and significantly improved recessional velocity accuracy for 22 known CSS found with 2dF.

We observed 110-120 targets in each of four 25-arcminute diameter fields (NE, SE, NW and SW) surrounding NGC 1399 -- the field coordinates, exposure times and target details are summarised in Table \ref{table:exposures}. 

\subsection{Target selection}
Our targets were selected from a SExtractor-derived \citep{Bertin..1996}\footnote{The Source Extractor software package is available from https://sourceforge.net/projects/sextractor.} catalogue obtained through deep imaging of the Fornax Cluster with the CTIO Mosaic Imager at the 4-m Blanco Telescope in Chile \citep{Karick..2005}. These images, covering an approximately $70^\prime\times70^\prime$ region centred on NGC 1399, are in the SDSS five-colour ($u^\prime g^\prime r^\prime i^\prime z^\prime$) wide-band photometric system. They have limiting magnitudes of $g^\prime\sim$25.1 and $r^\prime\sim$24.3, and a seeing-limited resolution between 0.8 and 1.24 arcsec ($\sim$70--110 pc at the Fornax Cluster distance). We selected point source targets based on their similarity to a point-spread function (i.e. having an $r^\prime$-band stellarity index $\geq 0.7$).

Subsequent to our observations, we obtained dereddened $g^\prime r^\prime i^\prime$ photometry \citep*{Karick..2007}, calibrated to SDSS DR4 using the extended Landolts stellar library \citep{Stetson..2000}. We use this photometry through the rest of our paper. The average colour offsets between the original and revised photometry for our targets across the four VLT fields are $g^\prime-r^\prime \simeq -0.17$ mag and $r^\prime-i^\prime \simeq +0.06$ mag.

\begin{figure}
\centering
\includegraphics[width=0.48\textwidth]{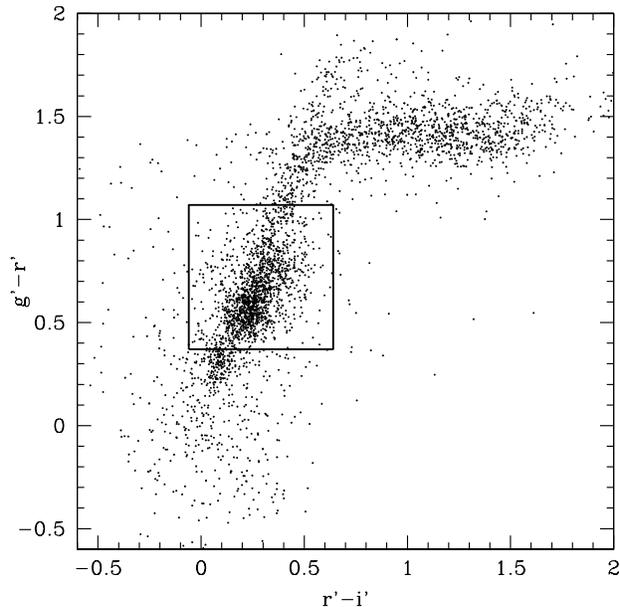}
\caption{Colour-colour plot ($r^\prime \leq 22.75$) of Fornax Cluster potential point source targets within the VLT-FLAMES field boundaries. Our VLT-FLAMES target selection box avoids the numerous red Galactic stars in a horizontal band at upper right.}
\label{fig:grritargets}
\end{figure}

% See SDSS color-color diagram explanation at http://cas.sdss.org/dr4/en/proj/advanced/color/making.asp
Fig. \ref{fig:grritargets} shows the distribution of point source targets in colour-colour space. CSS in the Fornax Cluster are generally found within or close to the well-defined locus of Galactic stars from lower-left to upper right, reflecting the dominant F-G-K main sequence spectral types of their stellar populations. We selected $\sim2100$ point source targets within a region of colour-colour space encompassing the known 2dF-derived CSS in Fornax ($0.37 < g^\prime-r^\prime < 1.07$ and $-0.06 < r^\prime-i^\prime < 0.64$ in the revised photometry). To compare our redshift results we specifically included several of the 2dF-derived CSS located within our field boundaries, and we prioritised the remaining targets by $r^\prime$ magnitude.

% This color-color plot is from /net/hubble0/firth/PAPERS/VLTFornaxPaper/analysis/VLTtargets.sm with singleplot = 1
% Subsequently, ramatched griz_nical and griz_MOSAIC targets within 1 arcsec and pasted amt and bmt into toto, then computed average gri offset between two sets of photometry (using nicalmosaic_diff.sm), enabling me to offset gr and ri target box for new photometry. Then updated VLTtargets.sm to read MOSAIC catalogue data and draw revised target colour box.

\subsection{Observations}
Our observations over two nights in 2004 November at Cerro Paranal, Chile were made under ESO observing programme 074.A-0508 with the FLAMES multi-fibre intermediate to high-resolution spectroscopic system attached to the Nasmyth-A focus of the 8-m Kueyen telescope. Each fibre has an aperture of 1.2 arcsec, requiring accurate target astrometry for reliable results -- since seeing during the observations was generally less than this aperture, the signal obtained was optimal (see Table \ref{table:exposures}).

% Seeing obtained from Raw file headers (using fv program) as key HIERARCH ESO TEL AMBI FWHM
% Field centres are from file /net/piscopia1/mjd/mirror/eso/notes/README
% bash command to get magnitude limit is similar to "sort -n +5 e1n1_results | tail"
\begin{table*}
\caption{VLT-FLAMES Observations}
\label{table:exposures}
\begin{tabular*}{1.0\textwidth}%
     {@{\extracolsep{\fill}}lccccccccc}
\hline
Field & RA & Dec & Targets & \multicolumn{2}{c}{Faint Limit} & \multicolumn{2}{c}{Blue 3964--4567\AA} & \multicolumn{2}{c}{Red 5015--5831\AA}\\
\cline{5-6} \cline{7-8} \cline{9-10}
Name & (J2000) & (J2000) & & $g^\prime$-band & $r^\prime$-band & Seeing & Exposure & Seeing & Exposure\\
 & \textit{h:m:s} & \textit{d:m:s} & & \textit{mag} & \textit{mag} & \textit{arcsec} & \textit{min} & \textit{arcsec} & \textit{min}\\
\hline
\\
NE & 03:39:40 & -35:12:00 & 114 & 23.6 & 22.5 & 0.5--0.8 & 100 & 0.9--1.9 & 100\\ % (e1n1)
SE & 03:39:30 & -35:36:45 & 119 & 23.5 & 22.3 & 0.5--0.9 & 103 & 1.4--3.1 & 115\\ % (e1s1)
NW & 03:37:17 & -35:15:27 & 118 & 23.1 & 22.4 & 0.5--0.9 & 105 & 0.7--1.0 & 105\\ % (w1n1)
SW & 03:37:15 & -35:38:30 & 117 & 23.4 & 22.4 & 0.6--1.2 & 120 & 0.5--0.7 & 120\\ % (w1s1)
\hline
\\
\end{tabular*}
\end{table*}

With no automatic dispersion correction (ADC), the VLT can efficiently observe only relatively narrow wavebands. We selected bands LR2 ($\lambda$ 3964--4567\AA, R 6400) and LR4 ($\lambda$ 5015--5831\AA, R 6000) covering several strong absorption lines, including the Balmer series (H$\beta$ 4861\AA, H$\gamma$ 4340\AA, H$\delta$ 4102\AA), CaII H and K (3968\AA\ and 3934\AA), the CH molecule or G band (4324\AA),the Mgb triplet (5167--5184\AA) and a number of FeI lines. We used the intermediate-resolution spectroscopy mode (600 lines mm$^{-1}$ and a blaze angle of 34$^{\circ}$) in order to maximise the number of targets observed at a S/N level sufficient to identify these key absorption lines. Resolution elements in the raw image are 0.094\AA\ (LR2 filter) and 0.136\AA\ (LR4 filter), but after sky subtraction and other pipeline processing this increases to 0.2\AA\ per bin. At this resolution a single absorption line feature can achieve a recessional velocity error of $cz\sim$15$ \; \mbox{km} \,\mbox{s}^{-1}$, but combining multiple features further reduces this error to a few kilometres per second.

\subsection{Data reduction and redshift measurement}
Raw spectral images were processed using the default settings with VLT-GIRAFFE pipeline data reduction software\footnote{http://www.eso.org/projects/dfs/dfs-shared/web/vlt/vlt-instrument-pipelines.html}. Sky-subtracted spectra were then examined with IRAF's \textsc{SPLOT} task to eliminate remaining sky features, cosmic rays, and CCD light leakage. To obtain redshifts we used IRAF's \textsc{XCSAO} to initially cross-correlate \citep{Tonry..1979} target spectra in each wavelength band to 1.4\AA-resolution standard stellar templates \citep*{Jacoby..1984}, identifying the VLT-FLAMES target spectra with the highest cross-correlation $R$-values. We then used these target spectra as cross-correlation templates, having the required resolution (0.2\AA) for higher precision cross-correlation to the remaining VLT-FLAMES target spectra.

We identified acceptable redshift measurements through cross-correlation $R$-values as follows

\begin{itemize}
\item In general, we selected the redshift result of the template with the highest $R$-value in either the blue or red waveband, using as a reliability threshold the minimum $R$-value of 4.2 for the 22 Fornax Cluster CSS originally discovered with 2dF and reconfirmed in our VLT-FLAMES results.
\item We rejected as unreliable cross-correlation results for targets where the highest $R$-value was less than 3.
\item Multiple independent wavelength bands are preferable to confirm the redshift of targets for which we only have low S/N, narrow waveband spectra. For targets with highest $R$-values between 3 and 4.2, we accepted those that produce similar redshifts for a majority of templates in both independent VLT-FLAMES wavebands (obtained on different nights). For the remainder, we individually inspected and assessed the strength of their absorption features in the blue and red wavebands. 
\end{itemize}

\subsection{Results}
\begin{figure}
\centering
\includegraphics[width=0.45\textwidth]{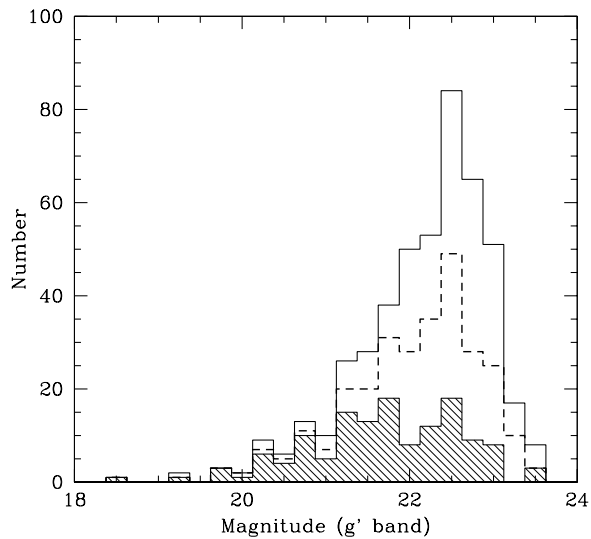}
\caption{Redshift measurement success of our VLT-FLAMES observations within the $g^\prime$-band range of our target catalogue. The upper line shows 468 observed targets in the four VLT-FLAMES fields. 289 measured redshifts with an $R$-value $\geq 3$ are shown by the dashed line, and 156 finally accepted redshifts are represented by the hatched region.}
\label{fig:obscompleteness}
\end{figure}

Fig. \ref{fig:obscompleteness} summarises our results by apparent magnitude -- we observed 468 targets from our colour-selected point source catalogue, and finally accepted 156 redshifts (from 289 redshifts with $R$-values $\geq 3$).  The accepted redshifts comprise 98 Galactic stars ($cz<600 \; \mbox{km} \,\mbox{s}^{-1}$), one background galaxy and 57 Fornax Cluster objects ($600 \leq cz \leq 2500 \; \mbox{km} \,\mbox{s}^{-1}$). Of 133 targets whose redshifts we rejected from individual inspection of their spectra, only 7 had (poorly correlated) redshifts within the Fornax Cluster range.

\begin{figure}
\centering
\includegraphics[width=0.45\textwidth]{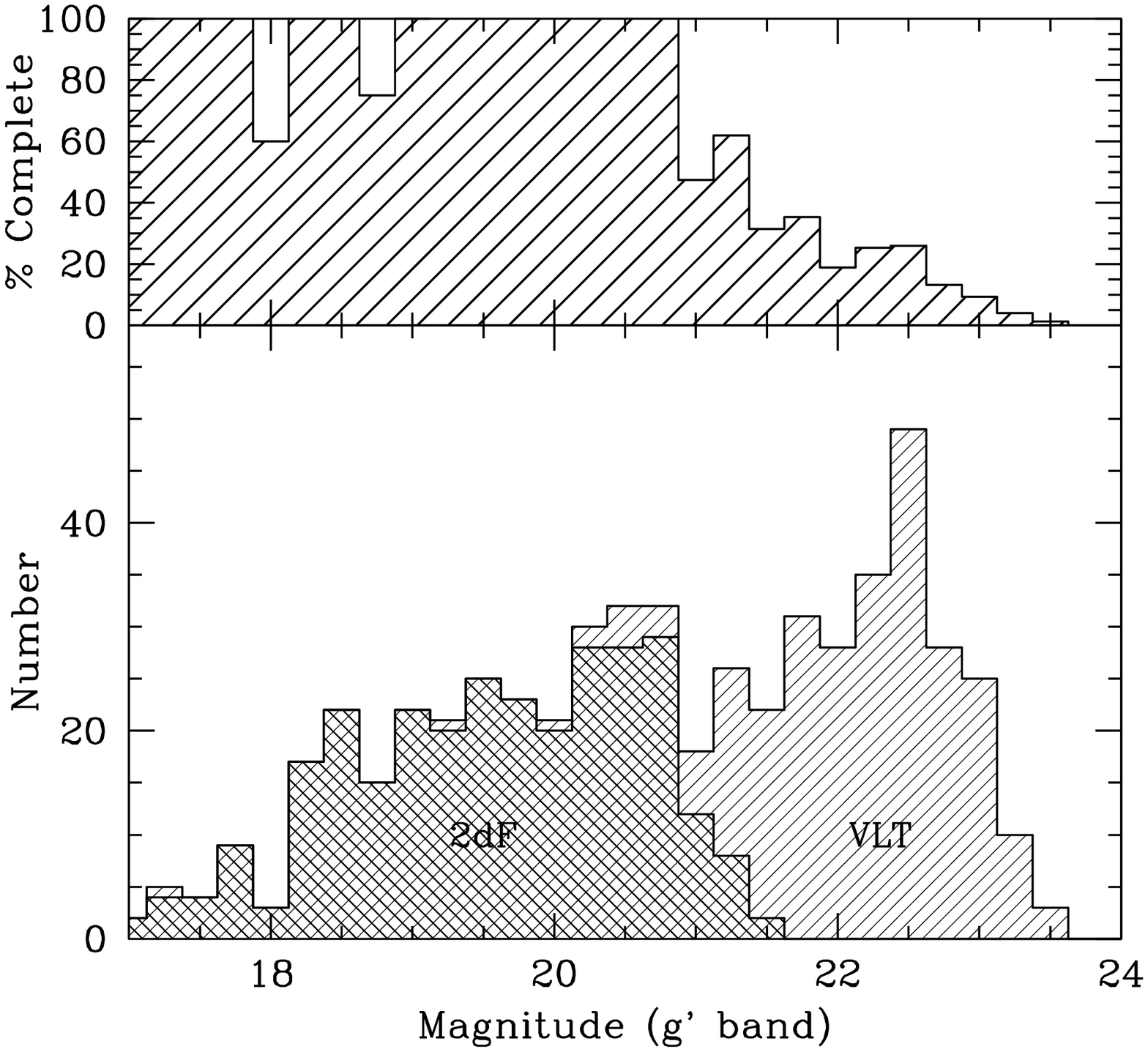}
\caption{Combined 2dF and VLT-FLAMES redshift completeness for colour-selected point sources in the four VLT fields. UPPER: The redshift information is essentially complete to $g^\prime\leq20.7$, and we can make reasonable inferences about the unknown redshifts to a faint limit of $g^\prime\sim23$. LOWER: Redshift information from 2dF observations has been significantly extended by our VLT-FLAMES results.}
\label{fig:redshiftcompleteness}
\end{figure}

The 57 confirmed Fornax Cluster objects comprise 4 known member galaxies (cross-checked with NED\footnote{The NASA/IPAC Extragalactic Database (NED) is operated by the Jet Propulsion Laboratory, California Institute of Technology, under contract with the National Aeronautics and Space Administration (http://nedwww.ipac.caltech.edu).} and inadvertently targeted due to point source confusion by SExtractor), and 23 known and 30 new CSS. As Fig. \ref{fig:redshiftcompleteness} shows, we have significantly extended the faint limit of known redshifts in our target catalogue.

% Summary table of targets observed and results - stars, UCDs, cluster galaxies, background galaxies.
% Table data derived from /net/hubble0/firth/PAPERS/VLTFornaxPaper/analysis/TargetData/resultsfile3.sxc using resultsfile.sm
% \begin{table*}
% \caption{VLT-FLAMES Observation Results}
% \label{table:results}
% \begin{tabular*}{1.0\textwidth}%
%      {@{\extracolsep{\fill}}lccc|cccc}
% \hline
% Magnitude & Targets & Redshifts & Accepted & Milky Way & Cluster & \multicolumn{2}{c}{Compact Stellar Systems}\\
% \cline{7-8}
% $g^\prime$-band & Observed & $R \geq 3$ & Redshifts & Stars & Galaxies & Reconfirmed & New\\
% \hline
% 17--18 & 1 & 1 & 1 & 1 & & & \\
% 18--19 & 1 & 1 & 1 &  & & &\\
% 19--20 & 5 & 4 & 4 &  & 1 & 3 &\\
% 20--21 & 30 & 25 & 21 & 7 & 1 & 17 & \\
% 21--22 & 102 & 77 & 50 & 35 & 2 & 2 & 12\\
% 22--23 & 249 & 137 & 51 & 48 & & & 3\\
% 23--24 & 44 & 22 & 8 & 7 & & & 1\\
% \hline
%  & & & & & & &\\
% \textbf{Totals} & 464 & 284 & 140 & 98 & 4 & 22 & 16\\
% \hline
% \end{tabular*}
% \end{table*} 

% The observational completeness adjustment figure and table are produced by VLTobscompleteness.sm

% The 2dF+VLT completeness adjustment figure and table are produced by VLTredshiftcompleteness.sm

\begin{figure}
\centering
\includegraphics[width=0.45\textwidth]{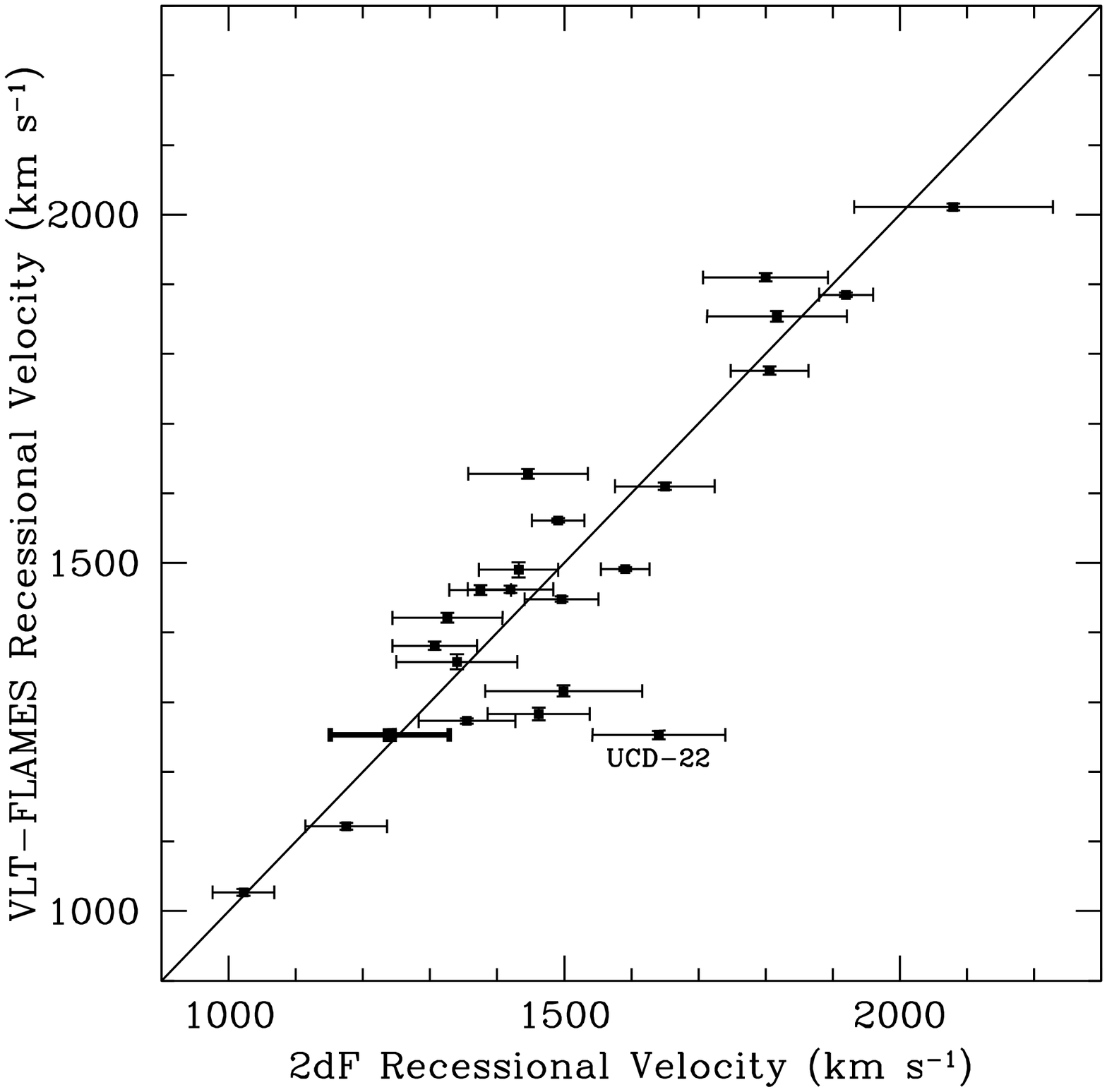}
\caption{Comparison of VLT-FLAMES recessional velocities for 22 reconfirmed CSS with those published by \citet{Drinkwater..2000b} and \citet{Gregg..2007}. $1\sigma$ VLT error bars are considerably smaller than the previous 2dF results. All 2dF data points are within $\sim2\sigma$ of the equality line, except UCD-22 (RA 03:37:43.49, Dec -35:15:10.2) for which however \citet{Mieske..2004} measured a recessional velocity (heavier error bar to the left) that is confirmed by our result.}
\label{fig:czcompare}
\end{figure}

Table \ref{table:VLTcatalog} lists positions, photometry and recessional velocities for the cluster dwarf galaxies and CSS observed in our four VLT-FLAMES fields, together with 2dF-derived CSS which are potentially bound to prominent galaxies surrounding NGC 1399 (see analysis in the next section). Fig. \ref{fig:czcompare} compares our VLT-FLAMES recessional velocity measurements with 2dF results by \citet{Drinkwater..2000b} and \citet{Gregg..2007}. The results are comparable (the VLT-FLAMES velocity mean for 22 re-observed CSS is $19.5 \; \mbox{km} \,\mbox{s}^{-1}$ lower than the 2dF velocity mean) except for the CSS at RA 03:37:43.49, Dec -35:15:10.2 (identified as UCD-22), for which however a \citet{Mieske..2004} measurement confirms our result. Our observations have reduced the average 2dF velocity error margin ($73 \; \mbox{km} \,\mbox{s}^{-1}$) to only $6 \; \mbox{km} \,\mbox{s}^{-1}$.

% Produced by czcompare.sm

\begin{table*}
\caption{Bound and Unbound CSS. Coordinates (J2000) and de-redenned photometry are from recalibrated CTIO imaging \citep{Karick..2007}. The stellarity index (sgc-r), derived by SExtractor from $r^\prime$ band photometry \citep{Karick..2005, Karick..2007}, compares the target light profile to a point-spread function (0 extended object; 1 point source). Velocities (cz) are heliocentric recessional. 2dF revised velocity data are derived from \citet{Drinkwater..2000b} and \citet{Gregg..2007}. FCOS data are from the Fornax Compact Object Survey \citep{Mieske..2004}. Bergond data are from VLT-FLAMES spectroscopy \citep{Bergond..2006}. $R$-values shown in brackets are from the previously published redshift measurement with the lowest error margin. The full version of this table can be found in the Supplementary Material section of the online paper at http://www.blackwell-synergy.com.}
\label{table:VLTcatalog}
{\scriptsize
\begin{tabular*}{1.00\textwidth}%
     {@{\extracolsep{\fill}}ccccccccccc@{}cl@{}c}
\hline
R.A. & Dec. & $g^\prime$ & $r^\prime$ & $i^\prime$ & sgc-r & 2dF & FCOS$^b$ & Bergond & \multicolumn{2}{c}{VLT-FLAMES} & & \multicolumn{2}{c}{Allocation}\\
\cline{10-11} \cline{13-14}
\textit{h  m  s} & \textit{d  m  s} & \textit{mag} & \textit{mag} & \textit{mag} &  & $km$ $s^{-1}$ & $km$ $s^{-1}$ & $km$ $s^{-1}$ & $km$ $s^{-1}$ & $R$-value. & & Galaxy & $[\Delta v] / v_{esc}$\\
(1) & (2) & (3) & (4) & (5) & (6) & (7) & (8) & (9) & (10) & (11)& & (12) & (13)\\
\hline
\\
\multicolumn{6}{l}{\textbf{Known Fornax Cluster dE Galaxies}}\\
03 36 37.25 & -35 23 09.4 & 20.87 & 20.34 & 20.12 & 0.38 & & & & 1401$\pm$13 & 3.6 & & FCC 171 &\\
03 37 17.91 & -35 41 57.6 & 20.55 & 19.93 & 19.65 & 0.00 & 1237$\pm$84 & & & 1279$\pm$13 & 3.7 & & FCC 194 &\\
03 39 13.30 & -35 22 17.2 & 19.20 & 18.54 & 18.20 & 0.03 & & 824$\pm$24 & & 794$\pm$6 & 9.9 & & FCC 222 &\\
03 40 23.54 & -35 16 36.5 & 20.52 & 19.89 & 19.60 & 0.03 & 2045$\pm$107 & & & 1814$\pm$11 & 4.0 & & FCC 241 &\\
\\
\multicolumn{6}{l}{\textbf{Known Fornax Cluster CSS}}\\
03 36 22.23 & -35 36 34.4 & 20.29 & 19.77 & 19.51 & 0.98 & 1462$\pm$76 & & & 1282$\pm$10 & 5.7 & & &\\
03 36 26.67 & -35 22 01.5 & 20.20 & 19.42 & 19.11 & 0.05$^c$ & 1499$\pm$117 & & & 1315$\pm$8 & 6.6 & & NGC 1381 & 4.1\\
03 36 27.69 & -35 14 14.0 & 20.12 & 19.27 & 18.90 & 0.95 & 1297$\pm$45 & & 1386$\pm$4 & & (29.6) & & NGC 1381 & 3.4\\
. & . & . & . & . & . & . & . & . & . & . & & . & .\\
. & . & . & . & . & . & . & . & . & . & . & & . & .\\
\\
\multicolumn{6}{l}{\textbf{New Fornax Cluster CSS}}\\
03 36 36.31 & -35 21 58.6 & 22.04 & 21.44 & 21.13 & 0.98 & & & & 1415$\pm$44 & 3.0 & & &\\
03 36 51.67 & -35 05 35.1 & 21.39 & 20.92 & 20.70 & 0.98 & & & & 1869$\pm$20 & 3.2 & & NGC 1380 & 0.1$^a$\\ 
03 36 54.57 & -35 39 26.9 & 21.39 & 20.79 & 20.52 & 0.98 & & & & 1469$\pm$10 & 5.1 & & &\\
. & . & . & . & . & . & . & . & . & . & . & & . & .\\
. & . & . & . & . & . & . & . & . & . & . & & . & .\\
\hline
\end{tabular*}

\begin{flushleft}
a. These CSS are considered to be gravitationally bound to a prominent cluster galaxy other than NGC 1399 - see analysis in the text.\\
b. FCOS catalogue identifiers (as sequenced in the table): 1-017, 2-0231, 2-078, 4-2028, 2-073, 0-2025, 3-2027, 1-2053, 1-060, 3-2004, 1-2083 and 3-2019.\\
c. Photometry of these CSS has been affected by partial blending with adjacent stars.\\
d. UCD3 is known to have an extended stellar halo (De Propris et al. 2005), and may be eventually classified as a dE,N galaxy.\\
e. Photometry of this CSS has been affected by blending with a background galaxy.
\end{flushleft}
}
\end{table*}

\subsection{Assessment of target selection criteria}
Having identified a significant number of spectroscopically-confirmed Fornax Cluster CSS, we can now assess the effectiveness of our target selection criteria. Table \ref{table:VLTcatalog} shows that nearly all the new, faint CSS are point sources with $r^\prime$-band stellarity index $\geq0.97$. CTIO images of the new systems confirm that they are unresolved point sources at 1 arcsec resolution, although as noted in Table \ref{table:VLTcatalog} the stellarity index values of 4 CSS have been reduced by blending with adjacent non-cluster objects.

\begin{figure}
\centering
\includegraphics[width=0.45\textwidth]{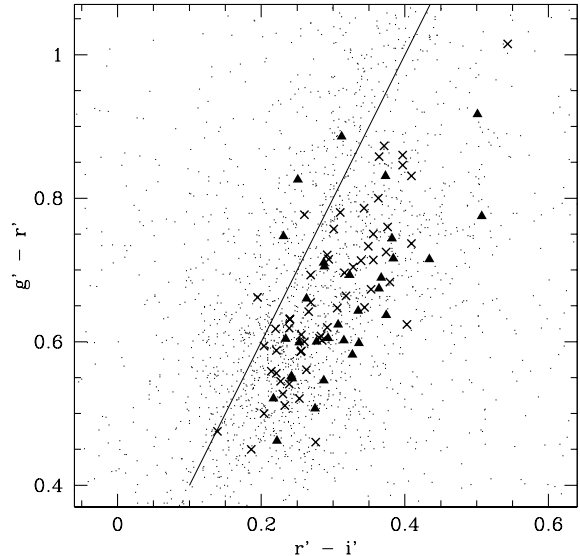}
\caption{Colour-colour plot of point sources (dots) in our four VLT-FLAMES fields, compared with the distribution of spectroscopically-confirmed Fornax Cluster CSS from 2dF (crosses) and from our VLT observations (triangles). The sloping line represents a potential future target selection criterion (discussed in the text) that would maximise our CSS yield.}
\label{fig:colourspace}
\end{figure}

Fig. \ref{fig:colourspace} compares the colour-space distribution of redshift confirmed CSS to the point source targets in our four VLT-FLAMES fields. A two-dimensional K-S test \citep{Numerical..1992} shows a statistically significant difference (K-S $D$-statistic 0.28, probability statistic\footnote{The probability statistic measures the likelihood that both samples are randomly selected from the same underlying population.} $<0.001$) between these two distributions -- spectroscopically-confirmed CSS generally have higher $r^\prime - i^\prime$ at given $g^\prime - r^\prime$ than the stellar locus.

% This color-color plot is from /net/hubble0/firth/PAPERS/VLTFornaxPaper/analysis/VLTcolorcolorplot.sm

Our redshift results have implications for future targeting of CSS in the Fornax Cluster central region. Of 156 accepted redshifts in our data, 4 were known galaxies and 22 were known Fornax UCDs, leaving 130 previously unknown redshifts from which our new faint CSS (31 objects listed in Table \ref{table:VLTcatalog}) represent a yield of $\sim25$ per cent. Based on our results, we could with hindsight have applied more restrictive target selection criteria to maximise our likelihood of finding the \textit{fainter} Fornax CSS by

\begin{enumerate}
\item Further restricting our point source catalogue to those having an $r^\prime$-band stellarity index $\geq0.97$. This would not apply to surveys for the more luminous CSS, which are sometimes more extended (with lower stellarity index values).
\item Limiting our targets to those having $g^\prime - r^\prime$ colours between 0.45 and 0.9 (see Fig. \ref{fig:colourspace}). 
\item Additionally limiting our targets to those with colour index $(g^\prime - r^\prime) \leq 2(r^\prime - i^\prime) + 0.2$, placing them to the right of the line shown in Fig. \ref{fig:colourspace} and excluding a significant number of targets located within the overall stellar locus (see Fig. \ref{fig:grritargets}).
\end{enumerate}

Although we would have missed targeting a small number of the newly-discovered CSS, we estimate that these tighter selection criteria would have significantly increased our faint CSS yield to $\sim45$ per cent of the accepted redshifts.

\section{Eliminating CSS bound to Galaxies surrounding NGC 1399}
The accumulated catalogue of radially-dispersed, redshift-confirmed CSS (2dF plus VLT-FLAMES) extends 47 arcminutes radially from NGC 1399, encompassing many prominent (non-dwarf) cluster galaxies within the Fornax Cluster central region. To analyse the distribution of CSS potentially associated with NGC 1399, we firstly eliminate systems that are satellites of (i.e. gravitationally dominated by) these prominent galaxies. Our approach is to provisionally associate CSS with a cluster galaxy if they are located within (or not far outside) its projected tidal radius -- determined by the opposing gravitational forces of the galaxy and NGC 1399.

\begin{figure}
\centering
\includegraphics[width=0.45\textwidth]{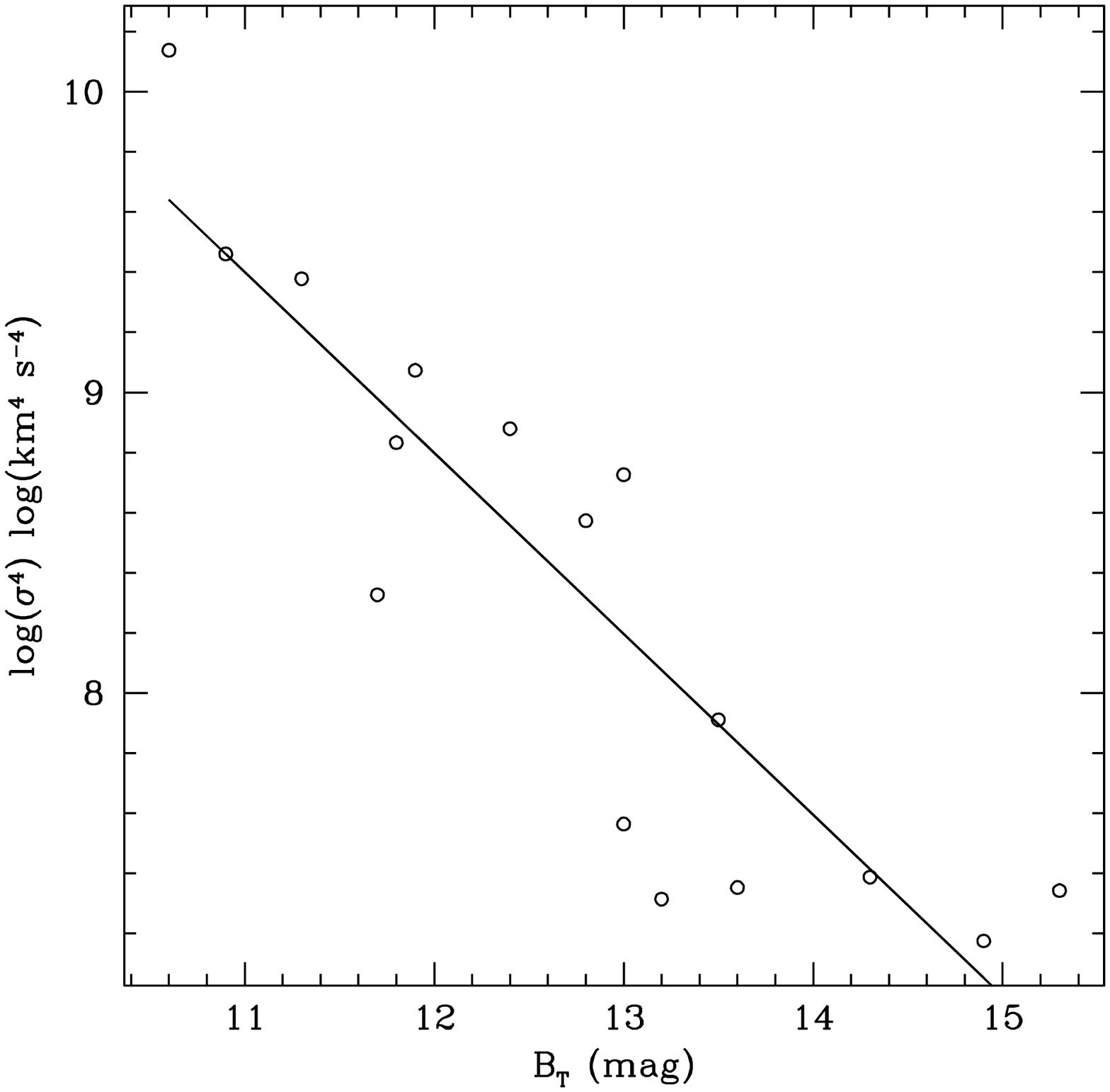}
\caption{Correlation of $B_T$ mag with measured central velocity dispersion ($\sigma$) for 16 prominent galaxies (open circles) surrounding NGC 1399. The correlation relation is shown as a continuous line.}
\label{fig:PromGalsBTveldisp}
\end{figure}

Table \ref{table:galaxies} lists the coordinates and observed features of the prominent Fornax Cluster galaxies within $1^\circ$ of NGC 1399, obtained from the Fornax Cluster Catalogue \citep{Ferguson..1989} and from HyperLEDA\footnote{We acknowledge the usage of the HyperLEDA database \citep{Paturel..2003}; http://leda.univ-lyon1.fr.}. Since HyperLEDA does not contain central velocity dispersion measurements for all these galaxies, we obtained a $B_T$ to $\sigma$ conversion relation ($log(\sigma^4)=[16.0\pm0.8]-[0.6\pm0.7]B_T$ based on the Faber-Jackson relation, which depends upon the virial theorem) from 15 elliptical or lenticular galaxies having $\sigma$ measurements, as shown in Fig. \ref{fig:PromGalsBTveldisp}. We then applied this conversion to obtain the estimated $\sigma$ values shown in italics in Table \ref{table:galaxies}. 

% This plot is from /net/hubble0/firth/PAPERS/VLTFornaxPaper/analysis/PromGalsBTveldisp.sm

Nearly all of the galaxies are elliptical or lenticular, for which we have calculated virial masses ($m_{vir}$) from their effective radii ($R_{eff}$ containing half the luminosity) and central stellar velocity dispersion ($\sigma$), using the formula $m_{vir}\simeq R_{eff}(1.65\sigma)^2/G$ \citep{Padmanabhan..2004}. For the two spiral galaxies FCC 121 and FCC 285, we have calculated dynamical masses ($m_{rot}$) from HyperLEDA data, using the formula $m_{rot} = R_{25}V_{max}^2/G$. $R_{25}$ is the radius along the projected major axis at the isophotal level $25 \, \mbox{mag} \; \mbox{arcsec}^{-2}$ in the B-band, and $V_{max}$ is the maximum rotation velocity corrected for inclination. To derive tidal radii we use the Jacobi limit approximation $r_t \simeq (m/3M)^{1/3}D$ \citep{Binney..1987} where $m$ is the mass of a prominent galaxy calculated as above, $M$ is the mass of NGC 1399 estimated from X-ray emission \citep{Jones..1997}, and $D$ is the projected separation. As expected, the average tidal radius is smaller for galaxies closer to NGC 1399.

\begin{table*}
\caption{Prominent Galaxies in the Fornax Cluster centre, showing published data together with our calculations of galaxy masses and tidal radii. FCC numbers, $B_T$ magnitudes and effective radii are from the Fornax Cluster Catalogue \citep{Ferguson..1989}. Central velocity dispersions ($\sigma$) are from HyperLEDA, unless (shown in italics) they are estimated as explained in the text. Coordinates (J2000) and heliocentric recessional velocities are from NED, except that we have obtained NGC 1399's recessional velocity from \citet{Richtler..2004}. Morphological types are from HyperLEDA.}
\centering
\label{table:galaxies}
{\scriptsize
\begin{tabular*}{1.00\textwidth}%
     {@{\extracolsep{\fill}}llcccccccccc}
\hline
\multicolumn{2}{c}{Catalogue Identifier} & $B_T$ & $R_{eff}$ & $\sigma$ & R.A. & Dec. & $cz$ & Type & Mass & \multicolumn{2}{c}{$r_t$}\\
\cline{1-2}\cline{11-12}
NGC & FCC & \textit{mag} & \textit{arcsec} & $km$ $s^{-1}$ & \textit{h:m:s} & \textit{d:m:s} & $km$ $s^{-1}$ & & $M_{sol} \times 10^{10}$ & \textit{kpc} & \textit{arcsec}\\
\hline
1399 & 213 & 10.6 & 37.2 & 342 & 03:38:29.08 & -35:27:02.7 & 1442$\pm$9 & E0 & $\sim$600$^a$ & & \\
\\
1365 & 121 & 10.2 & 130.0 & 151 & 03:33:36.37 & -36:08:25.4 & 1636$\pm$1 & Sb & 5.5 & 68 & 627 \\
PGC013230 & 136 & 14.8 & 16.3 & 153 & 03:34:29.48 & -35:32:47.0 & 1205$\pm$1 & S0 & 1.9 & 24 & 301 \\
1373 & 143 & 14.3 & 11.8 & 70 & 03:34:59.21 & -35:10:16.0 & 1334$\pm$2 & E & 0.3 & 14 & 156 \\
1374 & 147 & 11.9 & 24.0 & 186 & 03:35:16.59 & -35:13:34.5 & 1294$\pm$2 & E & 4.5 & 29 & 338 \\
1375 & 148 & 13.6 & 26.9 & 69 & 03:35:16.82 & -35:15:56.4 & 740$\pm$6 & S0 & 0.4 & 7 & 148 \\
IC0335 & 153 & 13.0 & 11.3 & 78 & 03:35:31.04 & -34:26:49.4 & 1619$\pm$6 & S0 & 0.5 & 29 & 270 \\
1379 & 161 & 11.7 & 20.9 & 121 & 03:36:03.95 & -35:26:28.3 & 1324$\pm$2 & E & 1.7 & 15 & 174 \\
1380 & 167 & 11.3 & 39.8 & 221 & 03:36:27.58 & -34:58:33.6 & 1877$\pm$12 & S0 & 15.4 & 58 & 464 \\
1381 & 170 & 13.0 & 12.9 & 152 & 03:36:31.68 & -35:17:42.7 & 1724$\pm$9 & S0 & 2.2 & 19 & 164 \\
1369 & 176 & 13.7 & 26.0 & \textit{88} & 03:36:45.25 & -36:15:22.4 & 1414$\pm$17 & S0-a & 3.5 & 37 & 394 \\
1380A & 177 & 13.2 & 12.6 & 67 & 03:36:47.49 & -34:44:22.6 & 1561$\pm$6 & S0 & 0.4 & 18 & 169 \\
1386 & 179 & 12.4 & 50.0 & 166 & 03:36:46.22 & -35:59:57.3 & 868$\pm$5 & S0-a & 5.0 & 19 & 329 \\
PGC013343 & 182 & 14.9 & 11.4 & 62 & 03:36:54.32 & -35:22:29.0 & 1657$\pm$19 & E-S0 & 0.3 & 7 & 66 \\
1387 & 184 & 12.3 & 50.1 & \textit{143} & 03:36:57.06 & -35:30:23.9 & 1302$\pm$12 & E-S0 & 6.2 & 15 & 172 \\
1380B & 190 & 13.5 & 16.3 & 95 & 03:37:08.96 & -35:11:42.1 & 1740$\pm$17 & E-S0 & 1.1 & 13 & 113 \\
1389 & 193 & 12.8 & 20.1 & 139 & 03:37:11.74 & -35:44:46.0 & 921$\pm$12 & E-S0 & 1.5 & 8 & 134 \\
1396 & 202 & 15.3 & 9.8 & 69 & 03:38:06.54 & -35:26:24.4 & 808$\pm$22 & E-S0 & 0.2 & 1 & 12 \\
ESO358-042 & 203 & 15.5 & 13.0 & \textit{47}& 03:38:09.15 & -34:31:06.7 & 1138$\pm$28 & S0 & 1.4 & 23 & 309 \\
PGC074808 & 207 & 15.9 & 8.5 & \textit{41} & 03:38:19.27 & -35:07:44.7 & 1420$\pm$20 & S0-a & 1.1 & 9 & 100 \\
PGC074806 & 208 & 17.3 & 11.7 & \textit{25} & 03:38:18.71 & -35:31:52.1 & 1720$\pm$50 & E-S0 & 1.9 & 4 & 32 \\
PGC074811 & 211 & 16.3 & 5.6 & \textit{36} & 03:38:21.52 & -35:15:34.1 & 2325$\pm$15 & E-S0 & 1.2 & 9 & 61 \\
1404 & 219 & 10.9 & 20.0 & 232 & 03:38:51.92 & -35:35:39.8 & 1947$\pm$4 & E & 8.8 & 13 & 100 \\
PGC013449 & 222 & 15.6 & 14.5 & \textit{45} & 03:39:13.36 & -35:22:18.6 & 824$\pm$24 & S0 & 1.1 & 3 & 52 \\
1427A & 235 & 13.4 & 36.3 & \textit{98} & 03:40:09.30 & -35:37:28.0 & 2028$\pm$1 & Irr & 7.0 & 29 & 215 \\
ESO358-051 & 263 & 14.6 & 20.5 & \textit{64} & 03:41:32.60 & -34:53:18.0 & 1724$\pm$6 & S0-a & 3.3 & 43 & 373 \\
PGC074933 & 274 & 16.5 & 12.0 & \textit{33} & 03:42:17.30 & -35:32:27.0 & 1073$\pm$76 & E-S0 & 1.2 & 18 & 246 \\
1427 & 276 & 11.8 & 32.4 & 162 & 03:42:19.43 & -35:23:33.5 & 1388$\pm$3 & E & 4.9 & 36 & 395 \\
1428 & 277 & 13.8 & 10.1 & \textit{85} & 03:42:22.73 & -35:09:14.4 & 1640$\pm$8 & E-S0 & 1.6 & 32 & 292 \\
ESO358-054 & 285 & 14.2 & 28.7 & - & 03:43:02.19 & -36:16:24.1 & 886$\pm$3 & SABd & 0.02 & 5 & 92 \\
\hline
\end{tabular*}
\begin{flushleft}
a. The mass of NGC 1399 is estimated from X-ray emission \citep{Jones..1997}.\\
\end{flushleft}
}
\end{table*}

For several of these galaxies, we can compare our tidal radius calculations to published GC radial distribution limits\footnote{GC radial distribution limits are defined to be where the projected radial profile of GC number density falls to the observed background level.} estimated from deep photometry. For NGC 1379 and NGC 1387, our tidal radii agree with the estimated GC limits of 162 and 186 arcsec respectively \citep*{Bassino..2006b}. For NGC 1374, NGC 1380 and NGC 1427, our tidal radii are significantly greater than the estimated GC limits of respectively 132 arcsec \citep{Bassino..2006b}, 250--300 arcsec \citep{Forte..2001} and 200 arcsec \citep{Kissler..1997} -- in these cases we suggest that the GC radial extent is not being tidally truncated by NGC 1399. For NGC 1404, which is within NGC 1399's X-ray halo, there are published GC limits of approximately 200 arcsec \citep{Richtler..1992} and 240 arcsec \citep{Forbes..1998} -- at least two times our calculated tidal radius. However, the allocation of bound CSS would not be altered if we doubled our calculated tidal radius for NGC 1404.

CSS within or near to the tidal radius of prominent galaxies other than NGC 1399 are indicated by entries in the last two columns of Table \ref{table:VLTcatalog}. We then apply an escape velocity filter to these provisionally identified bound CSS. In the absence of secondary distance indicators, redshift-derived recessional velocity is a combination of two unknowns -- peculiar velocity and distance. Since we have redshift information rather than true distances, we assume that a CSS is as close as possible in real space to its associated prominent galaxy, having a real spatial separation equal to the sky-projected separation -- this maximises the gravitational binding force and consequent escape velocity. A CSS and its associated galaxy are assumed to be gravitationally bound if their recessional velocities differ by less than the escape velocity $v_{esc} = (2GM_{galaxy}/r)^{1/2}$ at their projected radial spatial separation $r$. The final column of Table \ref{table:VLTcatalog} lists the ratio of the recessional velocity difference (between the CSS and its potentially associated galaxy) and the escape velocity at the projected radius. Therefore a ratio less than 1 implies they are gravitationally bound.

\begin{figure}
\centering
\includegraphics[width=0.5\textwidth]{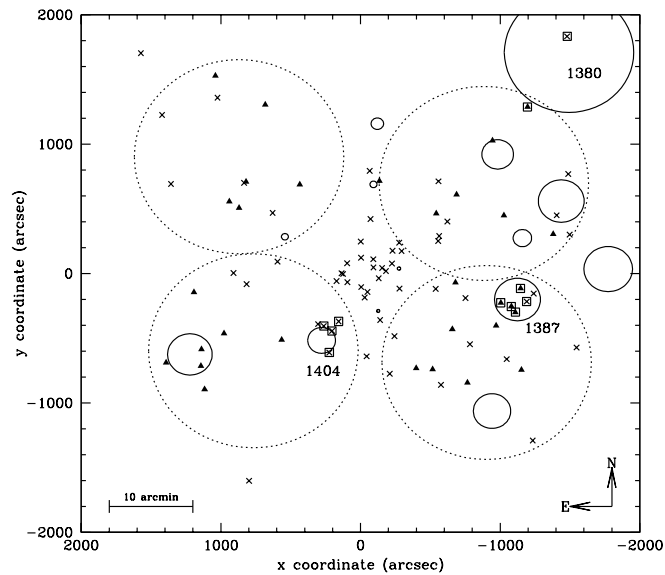}
\caption{CSS surrounding NGC 1399 (centre) from spectroscopy with 2dF \citep[crosses;][]{Drinkwater..2000b, Gregg..2007} and VLT-FLAMES (solid triangles), overlaid with our four 25-arcminute diameter VLT-FLAMES fields (dotted circles). We show the projected tidal radii (unfilled circles) of prominent cluster galaxies -- CSS considered to be gravitationally bound to these galaxies are enclosed within squares. Our wide-field spectroscopy provides new data between the redshift-confirmed \citet{Dirsch..2004} GCs which extend to 9 arcminutes from NGC 1399, and the wide-field photometry of \citet{Bassino..2006a} who estimates the extent of NGC 1399's GC system at $45\pm5$ arcminutes.}
\label{fig:fields}
\end{figure}

By this method we suggest that 11 of the redshift-confirmed CSS can be confidently associated with cluster galaxies other than NGC 1399, and we exclude them in our following analysis. Fig. \ref{fig:fields} shows the projected positions of redshift-confirmed CSS and prominent cluster galaxies surrounding NGC 1399, overlaid with the VLT-FLAMES field boundaries and the tidal radii of prominent galaxies.  Three prominent galaxies that have bound CSS are labelled.

% Sky plot in x y from VLT_rxyplot_UCDIGC.sm with plot parameter set to 1.
% Then superceded by TidalRadius_xyplot.sm to show tidal radii.

\section{Analysis and Discussion}
Having eliminated CSS bound to prominent cluster galaxies other than NGC 1399, we analyse the radial density distribution, recessional velocity distribution and colour-magnitude relation \textit{of the remaining unbound CSS}, using the combined data from 2dF \citep{Drinkwater..2000b, Gregg..2007} and VLT-FLAMES. We separate the unbound CSS into two sub-populations -- brighter or fainter than the metallicity and kinematic break found by \citet{Mieske..2004,Mieske..2006I} at $M_V = -11$.

We also compare the unbound CSS data to previously published data for NGC 1399's GC system and dwarf galaxies in the central region of the Fornax Cluster. For the spectroscopic surveys where individual object coordinates are available \citep{Dirsch..2004, Drinkwater..2000b, Bergond..2007, Gregg..2007}, we convert to CTIO $g^\prime r^\prime i^\prime$ photometry \citep{Karick..2007} by matching object positions in both catalogues. No coordinates were available for the 160 NGC 1399 GCs spectroscopically identified by \citet{Schuberth..2004}, so we were unable to compare our results in detail to this data. For candidate GCs derived from large sample photometric datasets, we use $g^\prime=V+0.3$ (where $20.3<g^\prime<25.7$) for the \citet{Dirsch..2003} survey, and $g^\prime=T_1-0.2$ (where $20.3<g^\prime<25.7$) for the \citet{Bassino..2006a} survey.

\subsection{Projected radial distribution}
% see VLTrotation.sm for calculation of rotation angle.
The projected distribution of unbound CSS (see Fig. \ref{fig:fields}) appears anisotropic with respect to NGC 1399 -- the major axis (best-fitting line through the projected distribution) of unbound CSS is rotated $84\pm5^{\circ}$ from N through E. This is not an artefact of the incomplete field coverage of our VLT-FLAMES observations, since the 2dF-derived CSS (covering the entire field shown in Fig. \ref{fig:fields}) displays the same projected distribution. 

Table \ref{table:density} tabulates the radial surface density distribution of unbound 2dF and VLT-FLAMES CSS located within the VLT field boundaries. The radial bins are set to contain approximately equal numbers of targets within our VLT-FLAMES fields, which extend radially between 4 and 34 arcminutes from NGC 1399. Completeness percentages compare the total number of accepted redshifts (from 2dF and VLT-FLAMES observations) to the total number of potential point sources targets, in each radial range of our VLT-FLAMES fields.

% Table of Radial Surface Density Distribution Calculations from VLTraddensity2.sm. Areas derived from skyarea.sm.
\begin{table}
\caption{Radial Surface Density Distribution of observed CSS around NGC 1399, with approximately equal numbers of CSS in each varying width radial bin.}
\centering
\label{table:density}
{\scriptsize
\begin{tabular}{@{}l@{\hspace{0.4cm}}c@{\hspace{0.4cm}}c@{\hspace{0.4cm}}c@{\hspace{0.4cm}}c@{\hspace{0.4cm}}c@{}}
\hline
Distance & Area in & Observed & Observed & Comple- & Estimated\\
from & VLT & Unbound & Number & teness & Number\\
NGC 1399 & Fields & CSS & Density &  & Density$^a$\\
\textit{arcmin} & \textit{$arcmin^2$} &  & \textit{$arcmin^{-2}$} & \textit{per cent} & \textit{$arcmin^{-2}$}\\
\hline
4.0-12.5 & 287 & 9 & 0.031 & 8.9 & 0.36$\pm$0.12\\
12.5-15.2 & 212 & 10 & 0.047 & 12.5 & 0.38$\pm$0.12\\
15.2-18.84 & 344 & 9 & 0.026 & 15.9 & 0.16$\pm$0.05\\
18.84-24.2 & 566 & 9 & 0.016 & 21.3 & 0.07$\pm$0.02\\
24.2-34.0 & 541 & 9 & 0.017 & 29.2 & 0.06$\pm$0.02\\
\hline
 & 1950 & 46 & & & \\
\hline
\end{tabular}
\begin{flushleft}
a. Uncertainties are based on Poisson statistics.\\
\end{flushleft}
}
\end{table}

\begin{figure}
\centering
\includegraphics[width=0.45\textwidth]{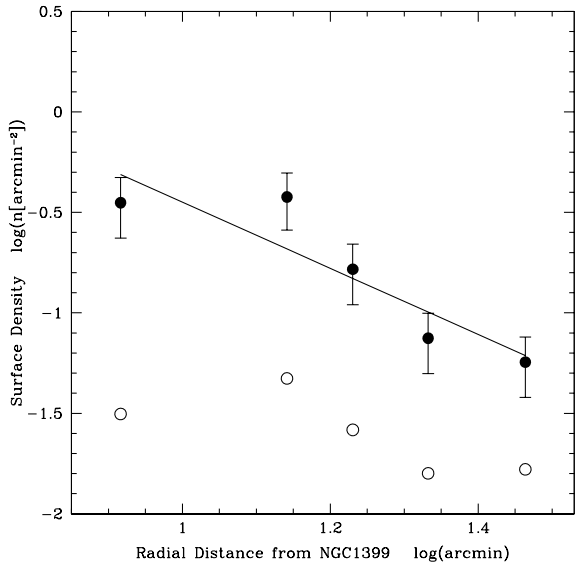}
\caption{Radial surface density distribution of unbound 2dF/VLT-derived CSS located within the VLT fields. The original data and completeness-adjusted estimates are shown as open and filled circles respectively. The best-fitting line for the completeness-adjusted data has a slope of $-1.6\pm0.2$.}
\label{fig:raddensity}
\end{figure}

In order to compare our unbound CSS radial density profile to previously published results by \citet{Bassino..2006a}, we used their combined blue and red GC number density data (Table 2 in their paper) between 9.5 and 28 arcmin from NGC 1399, as our VLT-FLAMES data only have significant areal coverage within these radial limits. Similarly, from Table 3 in \citet{Dirsch..2003} we used the combined blue and red GC number density between approximately 8 and 23 arcmin (0.94 and 1.36 in logarithmic units). Fig. \ref{fig:raddensity} shows that the best-fitting line to our completeness-adjusted surface density data ($18.6<g^\prime<22.9$) has a slope of $-1.6\pm0.2$ in log-log space. Within this radial range our result is similar to the slope of $-1.7$ \citep{Bassino..2006a} from photometry of candidate NGC 1399 GCs in a similar magnitude range ($19.8<g^\prime<22.8$), and the slope of $-1.5$ \citep{Dirsch..2003} from photometry of fainter candidate GCs ($23.3<g^\prime<25.7$).

% Radial surface density distribution of NGC1399 GC/IGC subset, from VLTraddensity2.sm.
% The projected area within 4 VLT fields at each radial distance bin is calculated by skyarea.sm

\subsection{Recessional velocity distribution}
\begin{figure}
\centering
\includegraphics[width=0.45\textwidth]{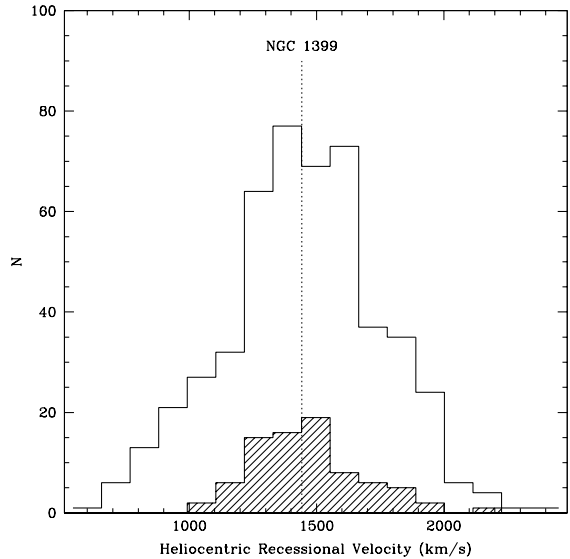}
\caption{The heliocentric recessional velocities of 2dF/VLT CSS not gravitationally bound to prominent galaxies surrounding NGC 1399 (hatched histogram) are compared with NGC 1399's inner GC system \citep{Dirsch..2004} and with NGC 1399 itself (vertical dotted line).}
\label{fig:velhist}
\end{figure}

In Table \ref{table:veldisp} we compare the heliocentric recessional velocity data from our observations with other published sources, and in Fig. \ref{fig:velhist} we show a velocity histogram comparing the unbound 2dF/VLT-FLAMES CSS to NGC 1399's inner GCs \citep[2 to 9 arcminutes from NGC 1399;][]{Dirsch..2004}.

% Table of Recessional Velocity Dispersions from VLTvelhist2.sm.
\begin{table}
\caption{Heliocentric Recessional Velocity Summary}
\centering
\label{table:veldisp}
{\scriptsize
\begin{tabular}{lcccc}
\hline
 & Source & N & Mean & Std. Devn.\\
 & & & \textit{$\mbox{km} \,\mbox{s}^{-1}$} & \textit{$\mbox{km} \,\mbox{s}^{-1}$}\\
\hline
NGC 1399 & \citet{Richtler..2004} & 1 &1442$\pm$9 & --\\
Inner GCs & \citet{Richtler..2004}$^a$ & 491 & 1445$\pm$14 & 303$\pm$10\\
Outer GCs & \citet{Schuberth..2004} & 160 & -- & $\sim276$\\
\\
Faint CSS & 2dF/VLT ($M_V \geq -11$) & 66 & 1464$\pm$28 & 225$\pm$20\\
Bright CSS & 2dF/VLT ($M_V<-11$) & 14 & 1464$\pm$61 & 229$\pm$45\\
All CSS & 2dF/VLT & 80 & 1464$\pm$25 & 224$\pm$16\\
\\
\parbox[t]{1.2cm}{Early-type Dwarfs} & various sources$^b$ & 88 & 1487$\pm$39 & 365$\pm$28\\
\hline
\end{tabular}
\begin{flushleft}
a. analysis based on \citet{Dirsch..2004} sample.\\
b. Early-type galaxies selected from FCC catalogue \citep{Ferguson..1989} with dwarf absolute magnitudes ($M_B>-17.6$) and with available recessional velocities. Sources for recessional velocities are as follows: \citet{Drinkwater..2000a, Drinkwater..2001, Mieske..2002, Karick..2005}\\
\end{flushleft}
}
\end{table}

The mean recessional velocity of the radially extended 2dF/VLT-FLAMES CSS differs by only $\sim20 \; \mbox{km} \,\mbox{s}^{-1}$ from that of NGC 1399's inner GC system. A \textit{t-test} shows this difference is marginally significant (50 per cent probability that they are drawn from the same population distribution). However, the CSS velocity dispersion is $\sim80 \; \mbox{km} \,\mbox{s}^{-1}$ less than that of the NGC 1399 inner GCs. An \textit{F-test} shows this difference to be highly significant, implying that the CSS and GC populations are kinematically different (0.1 per cent probability that they are drawn from the same population distribution). The CSS velocity dispersion is also $\sim50 \; \mbox{km} \,\mbox{s}^{-1}$ less than that reported by \citet{Schuberth..2004} for the more extended GC system (8 to 18 arcminutes from NGC 1399).

Similarly, there is only a small difference between the mean velocities of unbound CSS and early-type dwarf galaxies in the central region of the Fornax Cluster (63 per cent \textit{t-test} probability that they are from the same population), but a large difference in velocity dispersions ($<0.005$ per cent \textit{F-test} probability that they are from the same population).

The difference in mean recessional velocities of the bright ($M_V<-11$) and faint sub-populations of CSS is negligible, as confirmed by a \textit{t-test} ($>99$ per cent probability that they are from the same population).  The difference in their velocity dispersions is also negligible, as confirmed by an \textit{F-test} (86 per cent probability that they are from the same population).

% Rotation of GC/IGC System around NGC 1399. VLTrotation.sm
% \begin{figure}
% \centering
% \includegraphics[width=0.45\textwidth]{VLTrotation.eps}
% \caption{Relative heliocentric recessional velocity ($\delta cz$) of redshift-confirmed CSS measured from NGC 1399 along their best-fitting line. There is weak evidence for rotation of this structure ($61 \pm ?? \; \mbox{km} \,\mbox{s}^{-1}$.)}
% \label{fig:rotation}
% \end{figure}

% Fig. \ref{fig:posangle} shows the position angle and recessional velocity of redshift-confirmed CSS. Fit sinusoid to get position angle and amplitude of rotation??

% Rotation of GC/IGC System around NGC 1399. VLTrotation2.sm
% \begin{figure}
% \centering
% \includegraphics[width=0.45\textwidth]{VLTrotation2.eps}
% \caption{Position angle (measured clockwise from west) and recessional velocity of CSS, compared with that of NGC 1399 (dotted line).}
% \label{fig:posangle}
% \end{figure}

\begin{figure}
\centering
\includegraphics[width=0.45\textwidth]{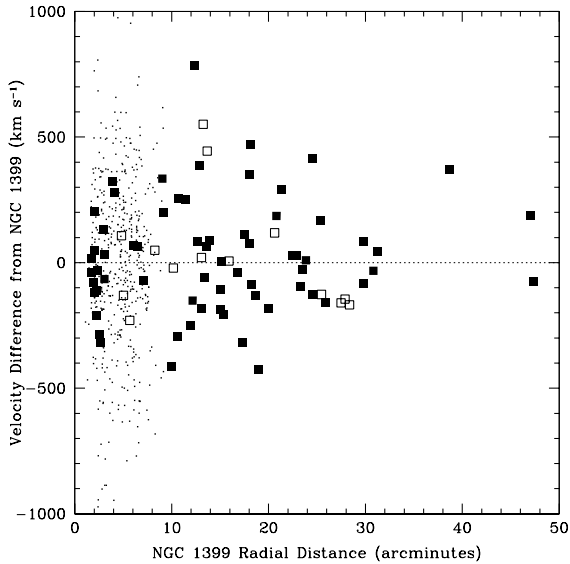}
\caption{Variation with radial distance from NGC 1399 in the heliocentric recessional velocity of bright ($M_V<-11$; open squares) and faint (filled squares) 2dF/VLT CSS, and of NGC 1399's inner GC system (\citet{Dirsch..2004}; small dots). The horizontal dotted line shows NGC 1399's recessional velocity.}
\label{fig:radiusvelocity}
\end{figure}

Fig. \ref{fig:radiusvelocity} compares the radial variation in heliocentric recessional velocity of bright and faint CSS with that of NGC 1399's inner GCs. The maximum radial extent of the bright CSS is $\sim28$ arcmin from NGC 1399, whereas the faint CSS extend to $\sim47$ arcmin.  No radial trend is apparent in these data for either sub-population.
% [confirms Mieske..2004 results ...??? Are there kinematically distinct sub-populations??].

We estimate the group rotation of unbound CSS around NGC 1399 by using a non-linear least squares fit of their position angles ($\theta_i$ with respect to NGC 1399) and recessional velocities ($v_i$). The fitted function is of the form $v_i = v_{sys} + A sin(\theta_i - \theta_0)$, where $v_{sys}$ is the mean velocity of the CSS, $A$ is the rotation amplitude and $\theta_0$ the position angle of the rotating system (measured from N through E). For the outermost blue GC clusters between 6 and $9 \; \mbox{arcmin}$ from NGC1399, \citet{Richtler..2004} found a rotation amplitude of $68\pm60 \; \mbox{km} \,\mbox{s}^{-1}$ and a position angle of $141^{\circ}\pm39^{\circ}$. With Monte-Carlo simulation, we find a mean unbound CSS recessional velocity of $1472\pm1 \; \mbox{km} \,\mbox{s}^{-1}$ (close to that of NGC 1399), rotation amplitude of $75\pm1 \; \mbox{km} \,\mbox{s}^{-1}$ \citep[similar to][]{Richtler..2004} and rotational position angle of $34^{\circ}\pm10^{\circ}$.

\subsection{Colour distribution}
\begin{figure}
\centering
\includegraphics[width=0.45\textwidth]{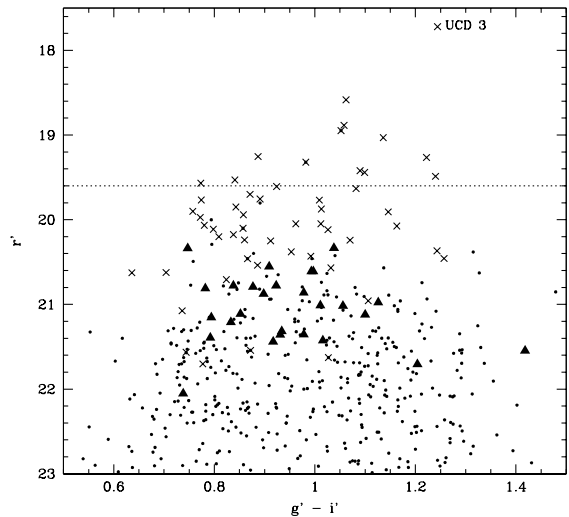}
 \caption{Colour-magnitude plot of redshift-confirmed unbound CSS in the Fornax galaxy cluster. Previously catalogued (crosses) and newly-discovered CSS (filled triangles) are compared with NGC 1399 GCs (small dots) from \citet{Dirsch..2004}. The dotted horizontal line divides the bright and faint CSS according to the metallicity break at $M_V=-11$ ($r^\prime=19.6$) found by \citet{Mieske..2006I}. UCD 3 is a known, particularly bright and extended object.}
\label{fig:colmag}
\end{figure}

Fig. \ref{fig:colmag} shows the colour-magnitude distribution of VLT/2dF-derived unbound CSS and NGC 1399's inner GC system. No colour bimodality is apparent or expected in our CSS data -- in the magnitude range $20<r^\prime<21$ \citet{Dirsch..2003} finds that GC colours are unimodal, and \citet{Bassino..2006a} has confirmed this for more radially extended CSS. The bimodal `red' and `blue' faint ($r^\prime\geq21$) GC sub-populations clearly found by \citet{Dirsch..2004} are also not discernible in our more radially extended faint CSS data, probably because our $g^\prime-i^\prime$ colour index is less sensitive to metallicity than their Washington $C-T1$ colour index. The `blue tilt' has not been previously detected in NGC 1399's GC system, and is also not apparent in our CSS data.

The mean colour index ($g^\prime-i^\prime=1.13\pm0.33$) of the 11 CSS bound to prominent galaxies (not shown in Fig. \ref{fig:colmag}) is redder than that of the 80 unbound CSS ($g^\prime-i^\prime=0.94\pm0.19$). This implied higher metallicity is expected for bound CSS which are closer to their parent galaxies \citep[for example][]{Brodie..2006}, although \citet{Barmby..2007} finds this correlation to be weak in galaxies of the Local Group. Excluding UCD 3 which may be a dwarf galaxy \citep{dePropris..2005}, the mean colour index of the bright unbound CSS ($g^\prime-i^\prime = 1.04\pm0.14\; \mbox{mag}$) is also slightly redder than that of the faint unbound CSS ($g^\prime-i^\prime=0.92\pm0.19 \; \mbox{mag}$), as might be expected given their higher metallicity \citep[see][]{Mieske..2006I}.

% This color-magnitude plot is produced by setting singleplot=2 in ../analysis/TargetData/colplot_allfields_plot.sm

\begin{figure}
\centering
\includegraphics[width=0.45\textwidth]{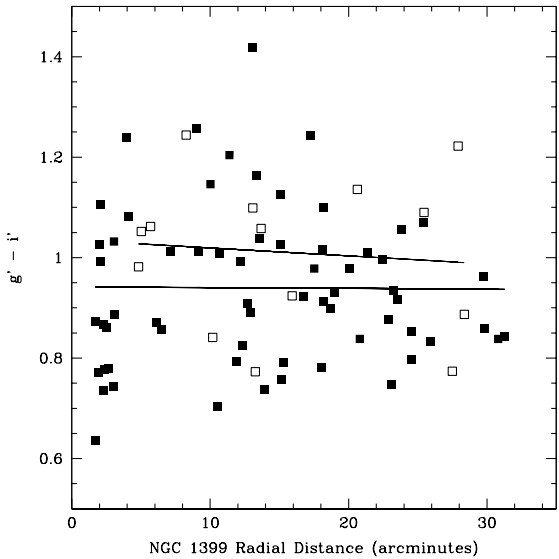}
\caption{We show $g^\prime - i^\prime$ colours of the bright ($M_V<-11$; open squares) and faint (filled squares) CSS as a function of radial distance from NGC 1399. The best-fitting lines for the bright (upper line) and faint (lower line) systems indicate insignificant radial colour change.}
\label{fig:radiuscolour}
\end{figure}

Fig. \ref{fig:radiuscolour} shows the $g^\prime - i^\prime$ colour index values of the bright and faint CSS as a function of radial distance from NGC 1399. There is no significant trend with increasing radius.

% This radius-colour plot is produced by ../analysis/VLTradiuscolour.sm

\section{Summary}
CSS are potential trace particles of galaxy assembly and observable fossil evidence of cluster evolution. The origin of bright CSS is unclear - they may be the bright tail of the GC distribution, merged superstellar clusters formed in galaxy interactions, or the tidally-stripped remnants of nucleated dwarf galaxies. The Fornax Cluster is a prime nearby target to systematically map the distribution of CSS over a large area with multi-fibre spectroscopy. While 2dF mapping \citep{Drinkwater..2000a, Gregg..2007} of the brighter radially-dispersed CSS has been completed within a $1^\circ$ radius around NGC 1399, extended exposure times or larger light-gathering power are needed to obtain reliable redshifts at the typical magnitudes of fainter CSS (see Fig. \ref{fig:redshiftcompleteness}).

\begin{figure}
\centering
\includegraphics[width=0.5\textwidth]{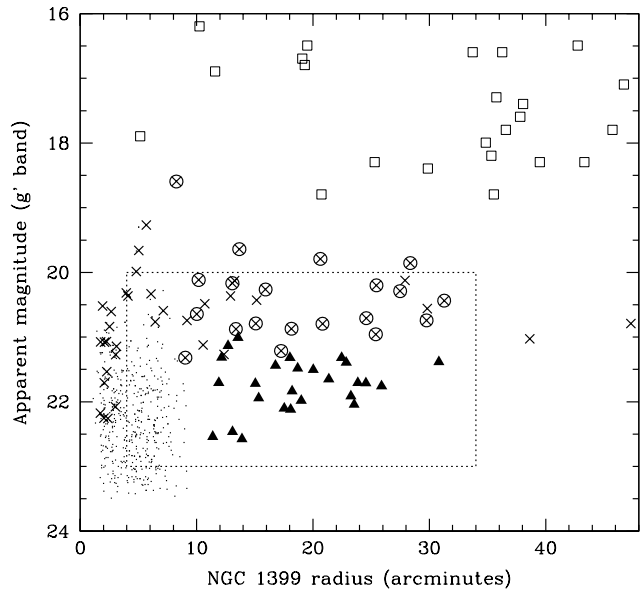}
\caption{The $g^\prime$-band magnitude and radial distance of redshift-confirmed CSS from NGC 1399. We show NGC 1399's inner GCs (dots) from \citet{Dirsch..2004}, and the more radially extended CSS from observations with 2dF (crosses) and VLT-FLAMES (solid triangles). CSS from 2dF observations which have been reconfirmed with VLT-FLAMES are circled. Early-type dwarf galaxies (open squares) are compiled from various sources as noted in Table \ref{table:veldisp}. The approximate limits of our VLT-FLAMES data are shown as a dotted line box -- 4 brighter 2dF-derived CSS were also targeted for redshift reconfirmation.}
\label{fig:radmagcontour}
\end{figure}

As Fig. \ref{fig:radmagcontour} illustrates, our VLT-FLAMES observations extend the faint limits of more widely dispersed CSS, and improve the recessional velocity accuracy of redshift-confirmed CSS in the central region of the Fornax Cluster. We have improved redshift accuracy for 23 known CSS and 3 member galaxies, obtained the first spectroscopic redshift measurement for member galaxy FCC 171, and measured redshifts for 30 new CSS. Our redshift results also establish significantly tighter target selection criteria, which we estimate could increase the yield of faint Fornax CSS to $\sim45$ per cent in future observations.

After excluding from our analysis 11 CSS dynamically associated with prominent galaxies surrounding NGC 1399, we are left with a catalogue of 80 redshift-confirmed unbound CSS surrounding NGC 1399. The following findings relate to these 80 unbound CSS (a sufficiently large data set to overcome small number statistics) unless otherwise stated

\begin{enumerate}
\item CSS are mostly located off the main stellar locus in colour-colour space, suggesting that they have a significant red stellar population component rather than being dominated by young blue stars.
\item The projected distribution of unbound CSS about NGC 1399 is anisotropic, following the galaxy distribution within the Cluster, and there is weak evidence for group rotation about NGC 1399.
\item The completeness-adjusted radial surface density profile of 46 spectroscopically-confirmed CSS located in our four VLT-FLAMES fields has a slope of $-1.7\pm0.2$ in log-log space. This result is similar to the slope of $-1.6$ \citep{Bassino..2006a} for photometry-derived candidate NGC 1399 GCs in a similar magnitude range, and the slope of $-1.5$ from \citet{Dirsch..2003} photometry of fainter candidate GCs. However, due to binning (of a smaller data set in this case), our result is subject to uncertainty from small number statistics.
\item The CSS mean heliocentric recessional velocity ($1464\pm25 \; \mbox{km} \,\mbox{s}^{-1}$) is between that of NGC 1399's inner GC system and that of the surrounding dwarf galaxies, but the CSS velocity dispersion is approximately $80 - 140 \; \mbox{km} \,\mbox{s}^{-1}$ lower than for these other cluster populations.
\item The mean colour of 14 bright CSS ($M_V<-11$) is 0.1 mag redder than the 66 faint CSS, suggesting they may have higher metallicity, subject to the uncertainty caused by small number statistics.
\item CSS show no significant trend in $g^\prime - i^\prime$ colour index with radial distance from NGC 1399.
\end{enumerate}

\section{Acknowledgements}
This work has been supported by grant No.~AST0407445 from the National Science Foundation. Part of the work reported here was done at the Institute of Geophysics
and Planetary Physics, under the auspices of the U.S. Department of Energy by Lawrence Livermore National Laboratory under contract No.~W-7405-Eng-48. Evstigneeva and Drinkwater acknowledge support from the Australian Research Council.

\bibliography{VLTFornax}

\label{lastpage}

\end{document}